\documentclass[11pt]{article}
\usepackage{amssymb,amsmath,graphicx,epsfig} 
\def\etal{{\em et al.}}
\def\issue(#1,#2,#3){{\bf #1}, #2 (#3)} 

\def\APP(#1,#2,#3){{\rm Acta Phys.\ Polon.} \ \issue({\bf #1},#2,#3)}
\def\ANP(#1,#2,#3){{\rm Annals of Physics} \ \issue({\bf #1},#2,#3)}
\def\ARNPS(#1,#2,#3){{\rm Ann.\ Rev.\ Nucl.\ Part.\ Sci.} \ \issue({\bf #1},#2,#3)}
\def\CPC(#1,#2,#3){{\rm Comp.\ Phys.\ Comm.} \ \issue({\bf #1},#2,#3)}
\def\CIP(#1,#2,#3){{\rm Comput.\ Phys.} \ \issue({\bf #1},#2,#3)}
\def\EPJ(#1,#2,#3){{\rm Eur.\ Phys.\ J.} \ \issue({\bf #1},#2,#3)}
\def\EPJD(#1,#2,#3){Eur.\ Phys.\ J. Direct\ C \ \issue({\bf #1},#2,#3)}
\def\IJMP(#1,#2,#3){{\rm Int.\ J.\ Mod.\ Phys.} \ \issue({\bf #1},#2,#3)}
\def\JHEP(#1,#2,#3){{\rm J.\ High Energy Physics} \ \issue({\bf #1},#2,#3)}
\def\JP(#1,#2,#3){{ J.\ Phys.} \ \issue({\bf #1},#2,#3)}
\def\MPL(#1,#2,#3){{Mod.\ Phys.\ Lett.} \ \issue({\bf #1},#2,#3)}
\def\NP(#1,#2,#3){{Nucl.\ Phys.} \ \issue({\bf #1},#2,#3)}
\def\NIM(#1,#2,#3){{ Nucl.\ Instrum.\ Meth.} \ \issue({\bf #1},#2,#3)}
\def\PL(#1,#2,#3){{ Phys.\ Lett.} \ \issue({\bf #1},#2,#3)}
\def\PR(#1,#2,#3){{ Phys.\ Rev.} \ \issue({\bf #1},#2,#3)}
\def\PRL(#1,#2,#3){{ Phys.\ Rev.\ Lett.} \ \issue({\bf #1},#2,#3)}
\def\SJNP(#1,#2,#3){{ Sov.\ J. Nucl.\ Phys.} \ \issue({\bf #1},#2,#3)}
\def\ZP(#1,#2,#3){{Zeit.\ Phys.} \ \issue({\bf #1},#2,#3)}
\def\be {\begin{equation}}
\def\ee {\end{equation}}
\def\bea {\begin{eqnarray}}
\def\eea {\end{eqnarray}}

\def\rpv {{R_p}\!\!\!/}
\def\bbbar{B^0-\overline{B}{}^0}
\begin{document}
\begin{center}
{\Large {\bf 
{$B\to\tau\nu$: Opening up the Charged Higgs Parameter Space with R-parity Violation}}}\\[5mm]
\bigskip
{Roshni Bose} \footnote{Electronic address: 123.roshni@gmail.com}
and {Anirban Kundu} \footnote{Electronic address: akphy@caluniv.ac.in}

\bigskip

{\footnotesize\rm 
Department of Physics, University of Calcutta, \\
92, Acharya Prafulla Chandra Road, Kolkata 700 009, India}

\normalsize
\vskip 10pt

{\large\bf Abstract}
\end{center}

\begin{quotation} \noindent 

The theoretically clean channel $B^+\to\tau^+\nu$ shows a close to $3\sigma$ discrepancy between 
the Standard Model prediction and the data. This in turn puts a strong constraint on the parameter space
of a two-Higgs doublet model, including R-parity conserving supersymmetry. The constraint is so strong 
that it almost smells of fine-tuning. We show how the parameter space opens up with the introduction of
suitable R-parity violating interactions, and release the tension between data and theory.

\vskip 10pt
PACS numbers: {\tt 12.60.Jv, 13.25.Hw}\\
\end{quotation}
\begin{flushleft}\today\end{flushleft}

\section{Introduction}
\label{sec1}

The purely leptonic decay $B^+\to\tau^+\nu$ has generated a lot of interest in recent times
because both BaBar and Belle collaborations found a sizable discrepancy with the Standard Model (SM)
prediction, which is quite clean and robust. The world average is 
\cite{ckmfitter-eps11} 
\be
Br(B\to\tau\nu) = (16.8\pm 3.1)\times 10^{-5}\,,
\ee
while the theoretical prediction is 
\be
Br(B\to\tau\nu)_{\rm SM} = \left(7.57^{+0.98}_{-0.61}\right)\times 10^{-5}\,,
\ee
which gives a tension at the level of $2.8\sigma$ 
\cite{ckmfitter-eps11}. 
The ratio of experimental and SM-expected  branching fraction
is approximately $2.22^{+0.50}_{-0.45}$.
The theoretical
uncertainty comes from the $B$ meson decay constant $f_B$ and the Cabibbo-Kobayashi-Maskawa
(CKM) matrix element $V_{ub}$. 

Using the Lattice QCD result of \cite{ckmlattice}
\be
f_{B_s} = (231\pm 3\pm 15)~{\rm MeV}\,,\ \ 
f_{B_s}/f_{B_d} = 1.235\pm 0.008\pm 0.033\,,
\ee
an SM-only explanation would require
\cite{ckmfitter-eps11} 
\be
\vert V_{ub}\vert = (5.10\pm 0.59)\times 10^{-3}\,,
\ee
which is clearly inconsistent with the indirect determination of $V_{ub}$ from the
sides of the Unitarity Triangle (UT)
\cite{ckmfitter-eps11}, 
\be
\vert V_{ub}\vert_{\rm indirect} = (3.49 \pm 0.13)\times 10^{-3}\,,
\ee
or the average of direct inclusive ($B\to X_u\ell\nu$) and exclusive ($B\to\pi
\ell\nu$) measurements \cite{taunu-babar},
\be
\vert V_{ub}\vert_{\rm measured} = (3.92\pm 0.09\pm 0.45)\times 10^{-3}\,.
\ee
This gives rise to the speculation that beyond-SM (BSM) physics might be at work here.
One of the possibilities is a possible new physics in $\bbbar$ mixing so that the
indirect measurement of $V_{ub}$ is not its SM value. Apart from that, the 
first candidate for such BSM physics is the charged Higgs boson $H^+$ of the
two-Higgs doublet models (2HDM) \cite{hou1993}, or of any supersymmetric model. The charged
Higgs contribution has a destructive interference with the SM $W$-mediated
contribution, so the solution comes only as a narrow band, centred roughly about
\be
m_H~({\rm GeV)} \approx 3.3\tan\beta\,,
\ee
where $\tan\beta$ is the usual ratio of the two vacuum expectation values \cite{akeroyd}.   
It has been argued that the solution looks rather unnatural and almost smells of
fine-tuning \cite{deschamps}. This solution, for the 2HDM model type II, also suffers 
serious tension from processes like $B\to D\ell\nu$, the ratio $Br(K\to\mu\nu)/
Br(\pi\to\mu\nu)$, $b\to s\gamma$, $Z\to b\bar{b}$, and the neutral B meson mass
differences $\Delta M_d$ and $\Delta M_s$. 
As was shown in \cite{deschamps}, the fine-tuned region disappears
when one takes all B-physics data into account at 95\% confidence limit (CL). 
However, it was recently 
shown in \cite{blankenburg} that a Minimal Flavour Violating
2HDM has a better agreement to these observables.

Models which embed the 2HDM, like supersymmetry (SUSY), have also been studied. The
conclusions, however, are not very enthusiastic \cite{akeroyd}. The reason is that in R-parity
conserving SUSY (the definition of R-parity is given later), which is
phenomenologically attractive because of its cold dark matter candidate, the 
SUSY effects to $B\to\tau\nu$ appear only as one-loop diagrams
with heavy electroweak gauginos and sleptons in the loop. Thus, the new amplitudes 
open up the parameter space only marginally \cite{akeroyd}.

R-parity violating (RPV) SUSY has also come up as another interesting option
\cite{akeroyd2,aida, cho}. 
The lepton and baryon numbers, L and B respectively, are good symmetries of
the Standard Model but {\em ad hoc} symmetries of 
the minimal supersymmetric SM, in the sense that one can write L and B violating
terms in the superpotential. However, conservation of both L and B
leads to a $Z_2$ symmetry called R-parity, and defined as
\be
R_p = (-1)^{{\rm 3B+L+2S} }\,,
\ee
where S is the spin of the particle. By definition, $R_p=+1$ for particles and
$R_p=-1$ for superparticles, and we demand $R_p$ to be a good symmetry of
the superpotential so that the $Z_2$ symmetry leads to a dark matter candidate.
On the other hand, if $R_p$ is not a good symmetry,
the signatures change drastically, because all superparticles, including the LSP,
can decay inside the detector. 

There can be 45 $R_p$-violating (RPV) couplings in the superpotential
coming from the renormalizable terms
\be
W_{\rpv} = \epsilon_{ab} \left( 
\frac12 \lambda_{ijk} L^a_i L^b_j \bar{E}_k + \lambda'_{ijk} L^a_i Q^b_j \bar{D}_k
\right)  + 
\frac12 \epsilon_{pqr} \lambda''_{ijk} \bar{U}^p_i \bar{D}^q_j \bar{D}^r_k\,,
\label{wrpv}
\ee
where $L$, $Q$, $E$, $U$ and $D$ stand for lepton doublet, quark doublet,
lepton singlet, up-type quark singlet, and down-type quark singlet superfields
respectively; $i,j,k$ are generation indices that can run from 1 to 3; $a,b=1,2$ 
are SU(2) indices, and $p,q,r=1,2,3$ are SU(3) indices; and 
$\lambda_{ijk}$ ($\lambda''_{ijk}$) are constructed to be antisymmetric in $i$ and $j$ ($j$
and $k$). The phenomenology of RPV supersymmetry, including the collider
signatures and bounds on these couplings, may be found in \cite{barbier}.
Apart from the trilinear terms, there can be bilinear R-parity violating
terms of the form of  $-\mu_i L_i H_2$, where
$H_2$ is the superfield that gives mass to charged leptons and down-type
quarks, in $W_{\rpv}$. We assume these bilinear terms to be zero
at the weak scale. This also relaxes the possible constraints
coming from the neutrino masses and mixing angles in presence of the bilinear
terms. However, even some trilinear combinations like $\lambda^{(')}_{ikl}
\lambda^{(')}_{jlk}$ can generate nonzero entries for the $ij$-th element of
the neutrino mass ${\cal M}_\nu$ \cite{barbier}. Putting the bilinear terms equal 
to zero means that we choose a particular basis in the $\{H_2, L_i\}$ space. 
The sneutrino vacuum expectation values need not be zero in this basis, but
that is not important for our case.

In this paper, we will try to use RPV SUSY from a different point of view. 
We will not constrain the RPV couplings from $B\to\tau\nu$; this has already
been done in \cite{aida,cho}, and there are other comparable or stricter
bounds \cite{dreiner06,tau3}. We will rather see how much the charged
Higgs parameter space in the $m_H$-$\tan\beta$ plane opens up because
of a new tree-level contribution coming from RPV SUSY. Considering the 
results of \cite{deschamps}, such a study is of serious importance. We will also
consider the possible effects of the complex phase of the RPV couplings.
It will be shown that with some couplings, the parameter space substantially
opens up and there is no longer any `fine-tuning'; with some other 
couplings, the effect is rather small because they are too tightly constrained.
Experimentally, this means that if the $B\to\tau\nu$ data remains 
anomalous, and we find the charged Higgs at some other point than 
that allowed by the narrow fine-tuning band, it will indicate another new
tree-level contribution; RPV SUSY is a prime candidate
for this.

The paper is arranged as follows. In Section II, we compile the relevant
formulae. The numerical analysis is taken up in Section III, and we 
summarize and conclude in Section IV.

\section{Relevant Expressions}

The decay width of $B\to\tau\nu_\tau$, in the SM, is given by 
\be
\Gamma (B\to\tau\nu_\tau) = \frac{1}{8\pi} G_F^2 \vert V_{ub}\vert^2 f_B^2 m_\tau^2 m_B \left(
1- \frac{m_\tau^2}{m_B^2}\right)^2\,,
\ee
which, in the presence of a charged Higgs, is modified by a multiplicative factor,
\be
\Gamma(B\to\tau\nu_\tau)_{\rm 2HDM} = \Gamma(B\to\tau\nu_\tau)_{\rm SM} \times
\left( 1- \frac{m_B^2}{m_H^2}\tan^2\beta\right)^2\,.
\ee

If R-parity is violated (there are two leptons in the final state, so we will consider only
L-violating interactions, {\em i.e.}, all $\lambda''$ couplings are assumed to be zero), there are
new contributions to the amplitude. These contributions, as has been pointed out in
\cite{aida}, can either be squark mediated (with a generic form of $\lambda'\lambda'$), or
slepton mediated (with a generic form of $\lambda\lambda'$). Each individual coupling can in
general be complex. While it is possible to absorb the phase of one coupling
by redefining the phase of the propagating sfermion, the phase of the second
coupling cannot be absorbed, and so in general this quadratic product of the RPV-couplings
is complex. 

The ATLAS and CMS experiments at the Large Hadron Collider have already ruled out
squarks below 800 GeV. Assuming a universal scalar mass $M_0$, one can safely
integrate out the propagating squark or slepton fields, and arrive at a four-fermion
interaction analogous to the Fermi interaction. As is the standard practice, we will
assume a hierarchical structure of the RPV couplings, so effectively only one 
product is nonzero at the weak scale. This might not be the case if the RPV couplings
are defined in the weak basis and one rotates them by a CKM-like mechanism
to get the relevant couplings in the mass basis \cite{dreiner2}, but the qualitative
results do not change much. Even in the mass basis, the RPV couplings $\lambda_{ijj}$
($i\not=j$) or ${\lambda'}_{ijj}$ are severely restricted, as they are possible sources
of neutrino Majorana mass terms.

One also notes that with RPV, the final state neutrino can have any flavour, depending
on the couplings. If it is a $\nu_\tau$, the $\rpv$-amplitude will add coherently with
the SM and 2HDM amplitudes; if it is $\nu_e$ or $\nu_\mu$, the addition will be
incoherent. For the latter case, the weak phase of the RPV coupling will
not matter.

The relevant four-fermion
operators for the decay $B\to\tau\nu$ may be obtained by 
integrating the sfermion field out in eq.\ 
(\ref{wrpv}). The expression reads \cite{dreiner06,tau3}
\bea
{\cal H}_{\rpv} &=& 
A_{jklm} (\bar\nu_j (1+\gamma_5) \ell_m) (\bar d_l (1-\gamma_5) u_k)
\nonumber\\
&& - \frac12 B_{jklm} (\bar\nu_j \gamma^\mu(1-\gamma_5) \ell_m) (\bar d_l \gamma_\mu(1-\gamma_5) u_k)\,,
\label{4fer}
\eea
where
\be
A_{jklm} = \sum_{i=1}^3 \frac{\lambda^\ast_{ijm}\lambda'_{ikl}}{4M_{\tilde\ell_{iL}}^2}\,,\ \ \
B_{jklm} = \sum_{i=1}^3 \frac{\lambda'_{mki}{\lambda'}^\ast_{jli}}{4M_{\tilde  d_{iR}}^2}\,.
\label{labc}
\ee
We take, keeping the recent ATLAS and CMS results in view, all sfermions to be 
degenerate at 1 TeV. 
The bounds scale with $M_{\tilde f}^2$, as is evident from
eqs.\ (\ref{4fer}) and (\ref{labc}). For our case, $k=1$, and $l=m=3$.

The contributions to the decay width of $B\to\tau\nu$ are given as
\bea
{\cal M}_{\rm {SM+2HDM}} &=& \frac{1}{\sqrt{2}}G_F V^\ast_{ub} 
\left( 1- \frac{m_B^2}{m_H^2}\tan^2\beta\right)\,,\nonumber\\
{\cal M}_{\rm squark} &=& \frac12 B_{j133}\,,\nonumber\\
{\cal M}_{\rm slepton} &=& -A_{j133}\times \frac{m_B^2}{m_\tau (m_b+m_u)}
\approx - 3.7 A_{j133}\,,
\eea
where we have used $m_B=5.27$ GeV, $m_\tau=1.777$ GeV, and the running mass
$m_b \equiv \overline{m_b}(\overline{m_b}) = 4.22$ GeV, corresponding to a pole mass
$m_b({\rm pole}) = 4.63$ GeV.

The branching fraction is 
\be
Br(B\to\tau\nu) = \frac{1}{4\pi} f_B^2 m_\tau^2 m_B \tau_B \left(
1- \frac{m_\tau^2}{m_B^2}\right)^2 \vert 
{\cal M}_{\rm {SM+2HDM}} + {\cal M}_{\rm squark/slepton} \vert^2\,,
\label{taunubr}
\ee
where $\tau_B = 1.525$ ps is the lifetime of $B^+$. Eq.\ (\ref{taunubr})
depends on the relative phase $\theta$ ($0\leq \theta< 2\pi$)
of $V^\ast_{ub}$, which by convention
is denoted by $\gamma$ or $\phi_3$, and the relevant RPV product coupling.
While eq.\ (\ref{taunubr}) is true only for $j=3$, {\em i.e.}, when the
emitted neutrino is $\nu_\tau$, for $j=1$ or 2 one can simply set
$\theta=\pi/2$.

Some of the relevant product couplings also contribute to other processes.
Most notable among them are the lepton flavour violating (LFV) decays $B_d\to
e^-\tau^+$ and $B_d\to \mu^-\tau^+$. They put a rather tight constraint
on the corresponding couplings. Other processes include the decay $B\to
\pi\nu\bar\nu$, from which the constraints are relatively poor, or LFV
decays of the top quark. The relevant expressions can be found in
\cite{dreiner06,tau3}. One can summarize the bound on the $\lambda\lambda'$
combination coming from $B_d\to \ell\bar{\ell'}$ decay as
\be
\vert\lambda\lambda'\vert = 0.017\left( \frac{Br(B\to\ell\bar{\ell'})}{25\times 10^{-6}}
\right)^{1/2} \left(\frac{\tilde M}{1000}\right)^2 
\left(\frac{m_b}{4.2}\right) \left(\frac{f_B}{0.2}
\right)^{-1}\,,
\label{b-ll}
\ee
where the sfermion mass $\tilde M$, the $b$-quark running mass $m_b$,
and the decay constant $f_B$ are all measured in GeV.

\section{Numerical Analysis}

\begin{figure}
\vspace{-10pt}
\centerline{
\rotatebox{-90}{\epsfxsize=5.5cm\epsfbox{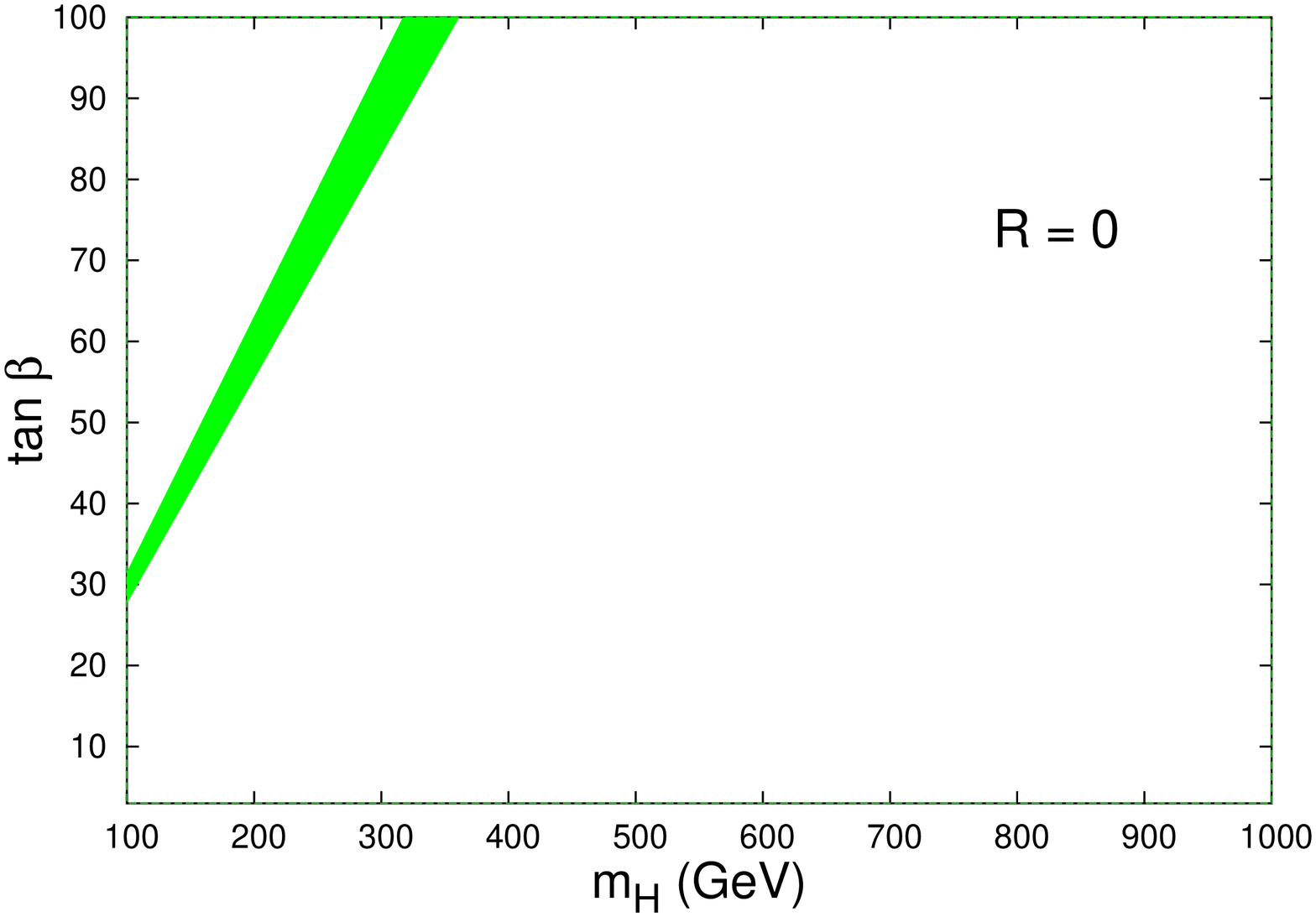}}
\rotatebox{-90}{\epsfxsize=5.5cm\epsfbox{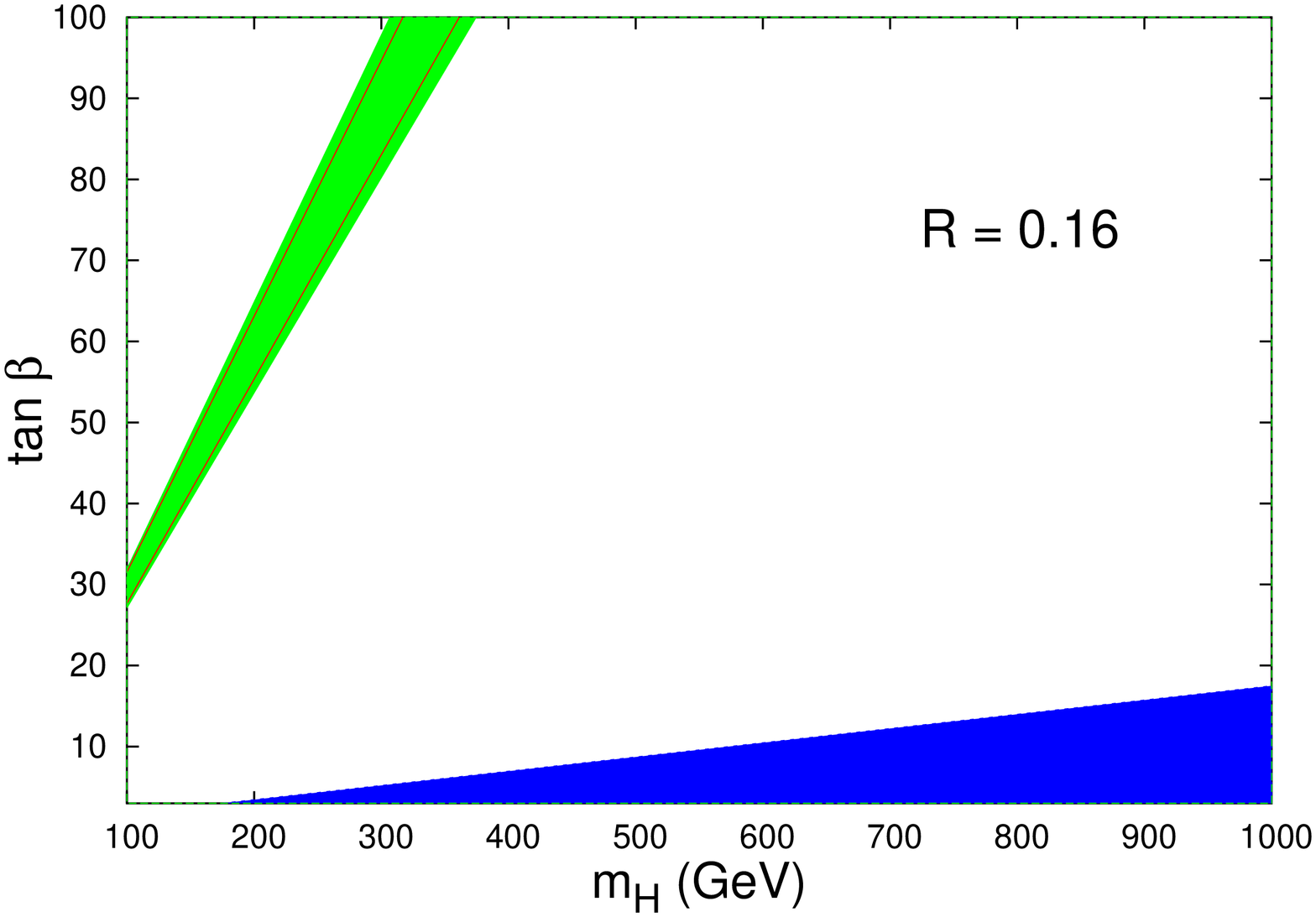}}
}
\centerline{
\rotatebox{-90}{\epsfxsize=5.5cm\epsfbox{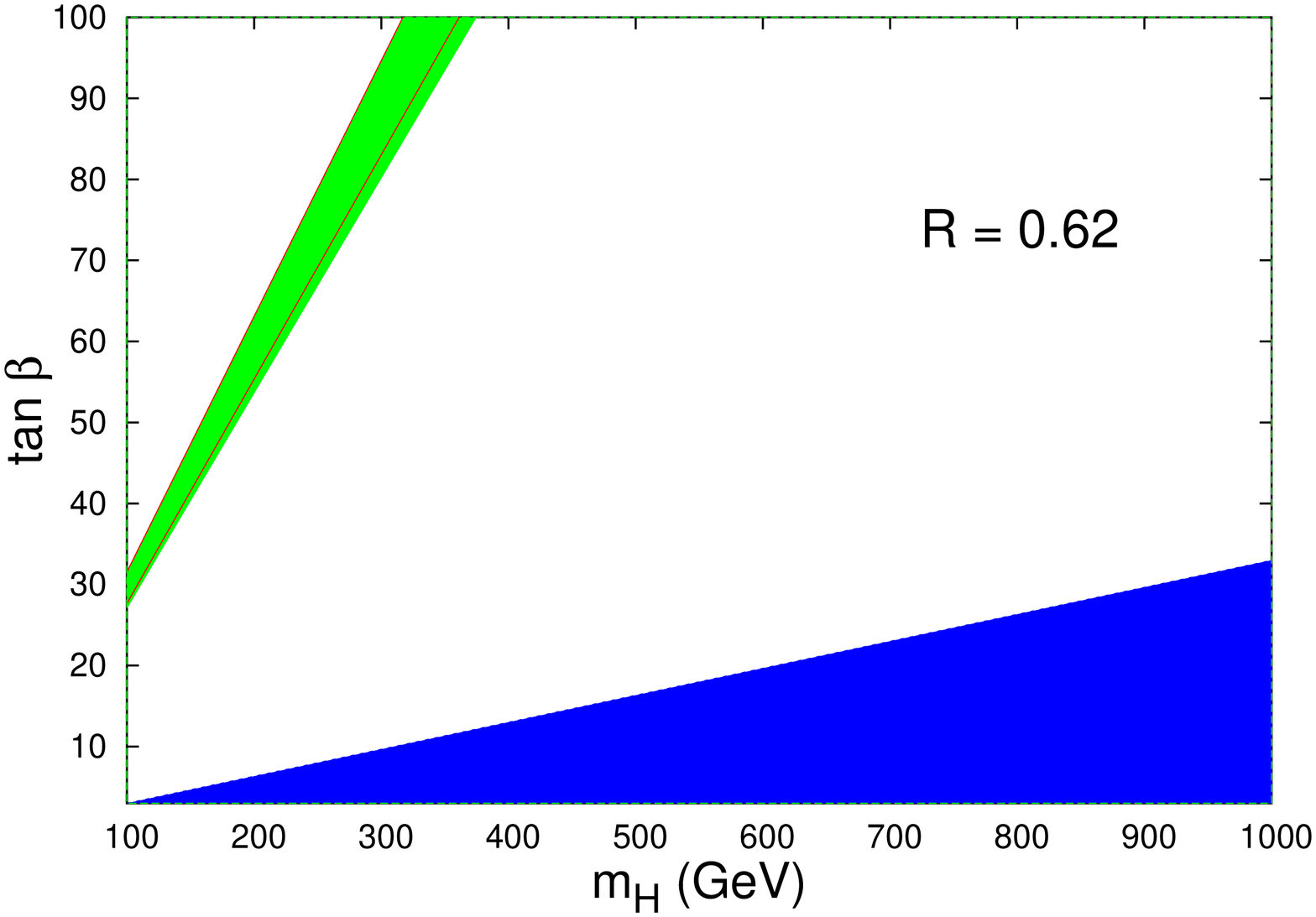}}
\rotatebox{-90}{\epsfxsize=5.5cm\epsfbox{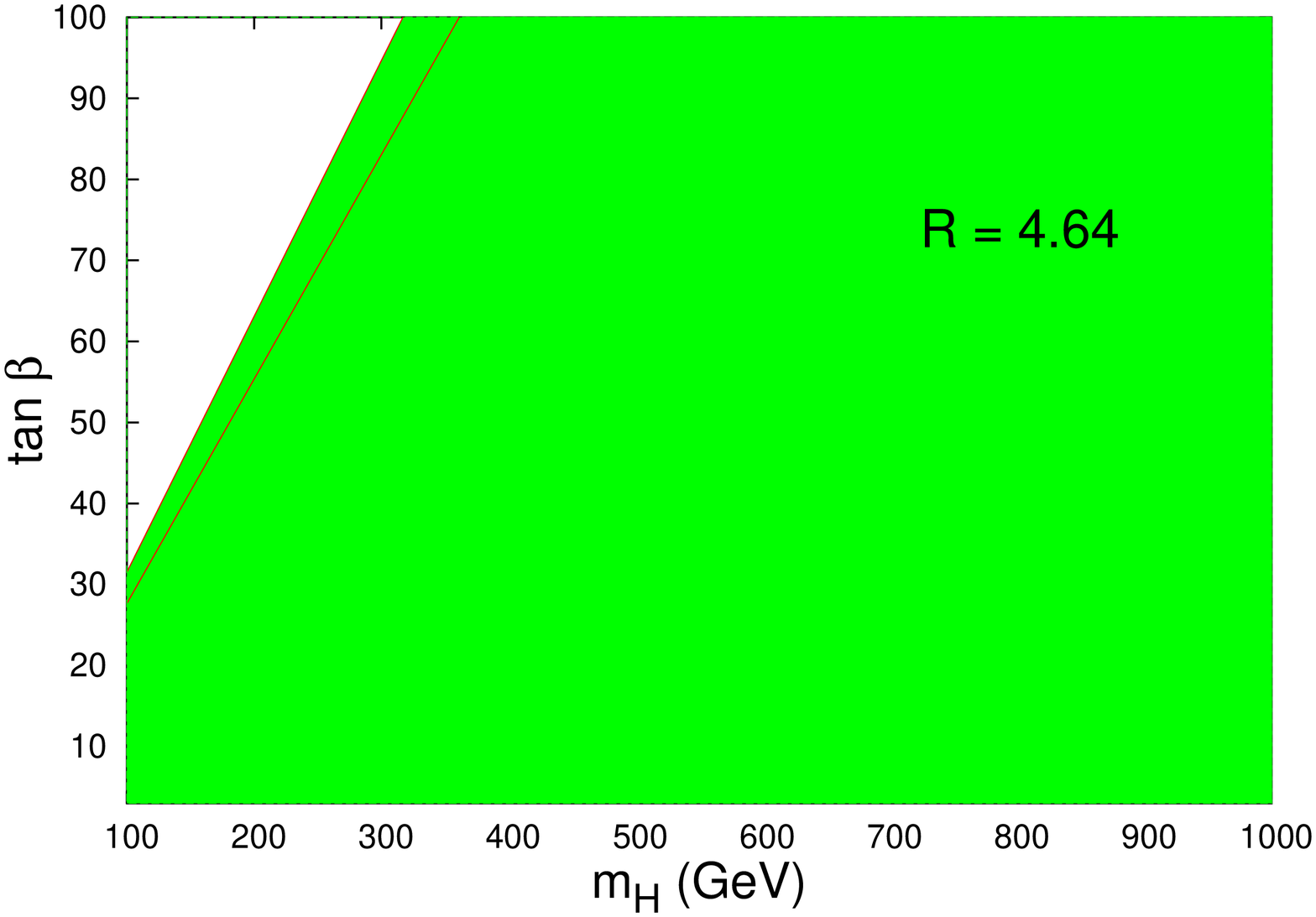}}
}
\caption{The allowed parameter space, at 95\% CL,
for the charged Higgs, as a function
of ${\cal R}$ defined in eq.\ (\ref{r-def}). Left upper plot shows the `only-2HDM'
fine-tuned parameter space. For the rest three plots, this region is that
between the solid lines. Right upper and left lower plots show the
allowed parameter spaces (blue/dark grey and green/light grey shaded regions) for different
values of ${\cal R}$. 
The lower right plot is for a large value of the RPV couplings, where the 
blue/dark grey and green/light grey regions merge.}
   \label{hspace}
\end{figure}

The upper bounds on the relevant RPV couplings $\lambda$ and $\lambda'$ are taken from \cite{barbier}, with
updates from \cite{kao-takeuchi}. Some of the product couplings are bounded by LFV decays of the
$B_d$ meson, and the numbers are taken from eq.\ (\ref{b-ll}) (also see \cite{dreiner06,tau3}). As all
these bounds are at $2\sigma$, we will show our numerical analysis at that confidence level only.

Neutrino masses put the tightest constraint on some of the RPV couplings. Assuming flat $\Lambda$CDM
cosmology, the total mass of all SM-like $\nu$ and $\bar\nu$ species is bounded \cite{wmap} to be
$\sum m_\nu < 0.56$ eV. Taking each entry of the neutrino mass matrix to be of the order
of 0.28 eV, and all sfermions degenerate at 1 TeV, one gets some typical bounds \cite{barbier}:
\be
|\lambda'_{i11}| < 5.3\times 10^{-1}\,,\ \ 
|\lambda'_{i33}| < 6.4\times 10^{-4}\,,\ \ 
|\lambda_{i33}| < 2.7\times 10^{-3}\,.
\ee
There are also single-coupling $2\sigma$ bounds \cite{kao-takeuchi},
\bea
&&|\lambda_{12k}| < 0.3\,,\ \ 
|\lambda_{13k}| < 0.3\,,\ \ 
|\lambda_{23k}| < 0.5\,,\nonumber \\ 
&& |\lambda'_{11k}| <  0.3\,,\ \ 
|\lambda'_{21k}| < 0.4\,,\ \ 
|\lambda'_{31k}| < 0.6\,,\ \ 
|\lambda'_{1k1} < 0.3\,,
\eea
where $k=1,2,3$. We need some more couplings on which the $2\sigma$ bounds at 1 TeV
are rather weak \cite{barbier},
\be
|\lambda'_{133}| < 1.8\,,\ \ 
|\lambda'_{2k1}| < 1.8\,,
\ee
and some more on which the limit comes from the expectation that they remain perturbative
at the weak scale. Since this is a matter of choice, we impose a flat cutoff at 2:
\be
|\lambda'_{132}|,~|\lambda'_{23k}|,~|\lambda'_{33k}| < 2\,.
\ee
The LFV decays of $B_d$ yield the following bounds:
\bea
|\lambda_{123}\lambda'_{113}|,~|\lambda_{233}\lambda'_{313}| < 0.017 &(B_d\to \mu^-\tau^+)\,,\nonumber \\ 
|\lambda_{123}\lambda'_{213}|,~|\lambda_{133}\lambda'_{313}| < 0.019 &(B_d\to e^-\tau^+)\,.
\eea
The mass difference for the $\bbbar$ system, $\Delta M_d$, gives \cite{jyoti3}
\be
|\lambda'_{i1j}\lambda'_{i3j}| < 0.04\,.
\ee
Table \ref{rpvbounds} summarizes the best bounds at the $2\sigma$ level.

\begin{table}[htbp]
\begin{center}
\begin{tabular}{||c|c|c||c|c|c||}
\hline
$\rpv$ coupling & Bound & ${\cal M}_{\rpv}/{\cal M}_{\rm SM}$&
$\rpv$ coupling & Bound & ${\cal M}_{\rpv}/{\cal M}_{\rm SM}$\\
\hline
$|\lambda'_{313}\lambda'_{i33}|$ ${}^\dag$   &   $3.8\times 10^{-4}$   &  $1.5\times 10^{-3}$  &
$|\lambda_{i23}\lambda'_{i13}|$    &   0.017   &   0.29 \\
$|\lambda'_{311}\lambda'_{131}|$   &   0.16   &  0.62  &
$|\lambda_{i13}\lambda'_{i13}|$    &   0.019   &   0.35 \\
$|\lambda'_{311}\lambda'_{231}|$   &   0.95   &  3.68  &
$|\lambda_{133}\lambda'_{113}|$ ${}^\ddag$    &   $8.1\times 10^{-4}$   &   0.023 \\
$|\lambda'_{311}\lambda'_{331}|$ ${}^\ddag$   &   0.04   &  0.16  &
$|\lambda_{233}\lambda'_{213}|$ ${}^\ddag$    &   $1.1\times 10^{-3}$   &   0.031 \\
$|\lambda'_{312}\lambda'_{332}|$ ${}^\ddag$   &   0.04   &  0.16  &
& & \\
$|\lambda'_{312}\lambda'_{1(2)32}|$    &   1.2   &  4.64  &
& & \\
\hline
\end{tabular}
\caption{Bounds on RPV couplings at $2\sigma$ with all sfermions degenerate
at 1 TeV. ${\cal M}_{\rm SM} = G_F \vert V^\ast_{ub}\vert/\sqrt{2}$. We have used 
a perturbative upper bound of 2 on some of the individual $\rpv$ couplings. 
The entries marked with $\ddag$ are to be added coherently with the SM 
amplitude, and those marked with $\dag$ only for $i=3$. 
For details, see text.}
\label{rpvbounds}
\end{center}
\end{table}


The opening up of the charged Higgs parameter space depends on the ratio
\be
{\cal R} = \frac{ {\cal M}_{\rpv} }{ {\cal M}_{\rm SM} }\,,
\label{r-def}
\ee
as displayed in Table \ref{rpvbounds}. For example, for ${\cal R} = 
1.5\times 10^{-3}$, the change is imperceptible. For other typical values,
we refer the reader to Fig.\ {\ref{hspace}}. 

The figure shows the charged Higgs
parameter space for $m_H\in [100:1000]$ GeV and $\tan\beta\in [3:100]$. 
For the coupling $\lambda'_{311}{\lambda'_{331}}^\ast$, the amplitude addition is 
coherent. There is a marginal enhancement on both sides of the pure-2HDM
band (only for those RPV amplitudes that add coherently with the SM one, the
enhancement is above the 2HDM band; for incoherent additions, only the lower
portion of the band might get allowed), 
and another region with low $\tan\beta$ opens up. This is the region
where charged Higgs contribution is insufficient to make up the deficit,
but that role is taken up by RPV SUSY. The lower left plot is for 
${\cal R} = 0.62$, where the relevant coupling is $|\lambda'_{311}\lambda'_{131}|$.
The emitted neutrino is $\nu_e$ and so the amplitudes add incoherently;
the parameter space opens up only on the lower side of the upper edge of
pure-2HDM region. Note that the gap between the two allowed regions
shrink. For ${\cal R}=4.64$, where the coupling is
either $|\lambda'_{312}\lambda'_{132}|$ or $|\lambda'_{312}\lambda'_{232}|$,
these two regions merge. The addition being incoherent, the allowed region
is again bounded by the upper edge of the 2HDM band.   

We do not show the plots for every possible coupling, because the trend
is obvious. The two regions merge at about ${\cal R}=1.18$, whether the sum is
coherent or incoherent. (However, no RPV couplings with such large ${\cal R}$ that can
add coherently with the SM amplitudes are allowed.)
Thus, the charged Higgs parameter space opens up
significantly, unless the corresponding RPV coupling is very tightly 
constrained. In particular, the low $\tan\beta$ region becomes allowed and 
alleviates the tension with other flavour observables \cite{deschamps}. 

One may ask what other processes are likely to be mediated by these couplings.
At the individual coupling level, a comprehensive list can be found in 
\cite{barbier}. At the product coupling level, all the $\lambda'\lambda'$ type products
can mediate $B_d\to\nu\bar\nu$, which might be observable at the next
generation B factories. Some of them can mediate four-quark interactions, like
$\lambda'_{312}\lambda'_{332}$ mediating $b\to s\bar{s}d$ (or $B_s\to\phi K_S$), but the
data is again inadequate to put further constraints. What we may emphasize is
that these channels are worth looking into. If lucky, one might even hope for
some LFV top decays too, like $t\to u \mu\tau$. 
In colliders like LHC, depending on the SUSY parameter space, one may observe
a stau decaying into jets. 

What happens if the RPV couplings are hierarchical not at the mass basis but
at the weak or flavour basis? In that case, one has to rotate these couplings
to the mass basis by some CKM-like mechanism, and this involves assumptions
about the mixing matrices in the right-chiral quark sector. However, a general
trend would be the appearance of the
complete set of all RPV couplings at the mass basis. With the neutrino bounds
at work, the constraints are expected to be much tighter, and hence less
allowed parameter space for the charged Higgs. The constraints are more
lenient if the mixing is in the up-quark sector, and our results do not
change much. If the mixing is in the down-quark sector, the constraints are
tighter, but one is still able to restore the low $\tan\beta$ region at
least for some of the product couplings.

\section{Summary and Conclusions}

The work was motivated by the fact that the tension between theory and
experiment for the decay width of $B\to\tau\nu$ requires at least another
tree-level contribution compatible in strength with the SM amplitude.
The most plausible candidate is a charged Higgs; however, the contribution
interferes destructively with the SM one, and one gets only a fine-tuned
region where the solution exists. Moreover, even this region is highly
disfavoured by other flavour observables. 

The next option is to use a model where the 2HDM is embedded, like 
supersymmetry. The R-parity conserving version has some one-loop corrections
to the $B\to\tau\nu$ amplitude, and that hardly helps alleviating the tension.
On the other hand, if one invokes R-parity violation, there are more
tree-level contributions to the decay, and the interference can be constructive.
Thus, the tension on the charged Higgs parameter space is relieved, and
one can have a massive charged Higgs at a sufficiently low value of
$\tan\beta$. One can also rephrase the conclusion: if a supersymmetric
charged Higgs is indeed found in this region, it will be worthwhile to look
for physics like R-parity violation, assuming the data on $B\to\tau\nu$ 
stands the test of time.
As a consistency check, we have made sure that the RPV couplings
satisfy all the existing bounds.

\centerline{\bf{Acknowledgement}}
RB acknowledges the University Grants Commission, 
Government of India, for financial support. 
The work of AK was supported by CSIR, Government of India, 
and the DRS programme of the University Grants Commission. 


  
\end{document}